\documentclass[11pt]{article}
\usepackage[normalem]{ulem}
\pdfoutput=1 

\usepackage{jheppub} 
\usepackage{url}
\usepackage{caption}
\usepackage{subcaption}
\usepackage{extarrows}

\usepackage[latin9]{inputenc}
\usepackage{float}
\usepackage{amsmath}
\usepackage{amssymb}
\usepackage{graphicx}
\usepackage{esint}
\usepackage{hyperref}
\usepackage{comment}
\usepackage{color}
\usepackage{microtype}
\usepackage{cleveref}
\usepackage{breakurl}
\usepackage{bbm}
\newcommand{\be}{\begin{equation}}
\newcommand{\ee}{\end{equation}}
\newcommand{\ben}{\begin{displaymath}}
\newcommand{\een}{\end{displaymath}}
\newcommand{\bea}{\begin{eqnarray}}
\newcommand{\eea}{\end{eqnarray}}
\def\K{K{\"a}hler }
   \newcommand{\rf}[1]{(\ref{#1})}
\newcommand{\vp}{\varphi}

\def\be{\begin{equation}}
\def\ee{\end{equation}}
\def\bea{\begin{eqnarray}}
\def\eea{\end{eqnarray}}
\def\ba{\begin{array}}
\def\ea{\end{array}}
\def\bit{\begin{itemize}}
\def\eit{\end{itemize}}

\def\a{\alpha}

\def\vp{\varphi}

 \makeatletter

\allowdisplaybreaks

\makeatother

\makeatletter
\DeclareRobustCommand{\rcite}[1]{%
  \rcite@aux#1,\@nil{#1}%
}
\def\rcite@aux#1,#2\@nil#3{%
  \if\relax#2\relax
    Ref.~\cite{#3}%
  \else
    Refs.~\cite{#3}%
  \fi
}
\makeatother

\hypersetup{
    colorlinks = true,
    citecolor = {blue},
    linkcolor = {blue},
    urlcolor = {blue},
}

\def\be{\begin{equation}}
\def\ee{\end{equation}}
\def\eqn#1{eq.~\eqref{#1}}

\def\rcite#1{ref.~\cite{#1}}

 \title{\Large \boldmath {\boldmath{Waterfall-modulated  $\alpha$-attractors}}}
\author[a]{Renata Kallosh,}
\author[a]{Andrei Linde,}
\author[b]{Yusuke Yamada}

\affiliation[a]{Leinweber Institute for Theoretical Physics at Stanford, 382 Via Pueblo, Stanford, CA 94305, USA}
\affiliation[b]{Cosmology, Gravity, and Astroparticle Physics Group, Center for Theoretical Physics of the Universe,
Institute for Basic Science (IBS), Daejeon, 34126, Korea}

\abstract{   Hybrid  $\alpha$-attractor models  \cite{Kallosh:2022ggf} can have significantly greater values of $n_{s}$ and smaller $r$, while preserving the relation $r\cong 3\a (1-n_s)^2$, which is valid for exponential  T- and E-models at large values of the inflaton field. Here we study single-field $\a$-attractors with features inspired by hybrid models: one can uplift the potential, and one can also have a waterfall regime that leads to a premature termination of inflation near the critical point $\vp_c$.  This allows one to increase the effective number of e-foldings $N_c$ in formulas like $n_s\simeq 1-{2\over N_c}$, $r\simeq {12 \alpha\over N_c^2}$. By changing the waterfall's steepness and location, one can continuously move the predictions along the curves with $r\cong 3\a (1-n_s)^2$ as $n_s$ increases and $r$ decreases.  We also study the effect of waterfall insertions and uplift on $n_s$ in quintessential $\alpha$-attractors that describe inflation and dynamical dark energy. 
 }

\begin{document}

\maketitle

 \parskip 5pt

\section{Introduction}\label{Sec:1}

The trademark property of the large family of $\alpha$-attractors introduced in  \cite{Kallosh:2013yoa} was the stability of the cosmological predictions of these models with respect to significant changes of their potentials: 
\be
\label{pred}
n_{s} \simeq 1-{2\over N_*} \ , \qquad r \simeq {12\alpha\over N_*^{2}} \ .
\ee 
where $N_*$ is the number of e-folds between the horizon exit of the pivot scale and the end of inflation.  

The value of $n_{s}$ in these models does not depend on any model parameters and coincides with the predictions of the Starobinsky and Higgs inflation models. The prediction of $r$ also does not depend on the potential; it is determined only by the \K curvature of the moduli space $\mathcal{R}_K= -{2\over 3\a}$.  

All models discussed in \cite{Kallosh:2013yoa} have a universal property. Their potentials have a minimum at $V = 0$ and a plateau, which is approached exponentially fast,
\be\label{aplateau}
V= V_{0}\left(1-c \, e^{-\sqrt{2\over 3\alpha }\vp}+ ... \right) \ .
\ee 
Here $c$ is a constant, which can be absorbed into a redefinition of the canonically normalized field $\vp$.  Using this expansion, one can derive a general relation between $n_{s}$ and $r$ which is valid at $\vp \gg  \sqrt{6\alpha}$: 
\be
 r\simeq  3\a(1-n_s)^2 \ .
\label{srr}
\ee
For a while, it seemed very difficult to deviate significantly from these predictions. However, one can avoid this rule if, for example, the last O(50) e-foldings of inflation occur at $\vp \leq  \sqrt{6\alpha}$, where an expansion \rf{aplateau} does not work \cite{Kallosh:2026tke}. A more radical way to do it was found in the context of hybrid $\alpha$-attractors \cite{Kallosh:2022ggf,Braglia:2022phb}.

The basic idea is very simple. In the slow-roll approximation,
\be\label{nsslow}
n_{s} = 1 -3\big(V'/V\big)^{2}+ 2 V''/V \ .
\ee
In hybrid inflation \cite{Linde:1991km,Linde:1993cn}, the potential of the inflaton field $\vp$ is uplifted by the Higgs-type potential of a second field $\chi$:\, $V(\vp) \to V(\vp)+ V_{\rm up}(\chi) + ...$. This increases the value of the potential during inflation driven by the field $\vp$ without changing its derivatives with respect to $\vp$. According to \rf{nsslow}, this fact alone is sufficient to increase $n_{s}$. 

In addition, inflation in this scenario ends not at the time when the slow roll of the field $\vp$ ends, but earlier, at greater values of $\vp \sim \vp_{c}$, when a ``waterfall'' instability with respect to the generation of the field $\chi$ triggers the end of inflation. At large $\vp_{c}$,  the uplifted $\alpha$-attractor potential \rf{aplateau}  approaches a constant along the valley $\chi = 0$, whereas its derivatives with respect to $\vp$ become exponentially small. This makes the last two terms in \rf{nsslow} very small,  $3\big(V'/V\big)^{2} \ll 2 \Big|V''/V\Big|\ll 1$.  

That is why a combination of an uplift of the $\alpha$-attractor potential and a premature end of inflation can increase $n_{s}$ all the way to $n_{s} = 1$. This should not come as a surprise, since the original (non-attractor) hybrid inflation models \cite{Linde:1991km,Linde:1993cn} typically predicted $n_{s}$ to be very close to $n_{s}=1$. In the case of $\alpha$-attractors, an increase of $n_{s}$ is easily controllable by the choice of $V_{\rm up}$ and $\vp_{c}$. For $V_{\rm up} = 0$ and $\vp_{c}= 0$ the predictions for $n_{s}$ and $r$ return to \rf{pred}.

Note that the uplifted potential of the field $\vp$ still has a plateau that can be described by \rf{aplateau}, albeit with a greater value of $V_{0}$ due to the uplift. Therefore, the universal relation \rf{srr} remains intact as long as the perturbations are produced at $\vp \gg \sqrt{6\alpha}$. And this means that an increase of $n_{s}$ is accompanied by a corresponding decrease of $r$. In fact, one can show that the standard expression \rf{pred} also remains valid if one replaces $N_{*} = O(50)$ in \rf{pred} by a phenomenological parameter $N_{c} $ depending on the detailed structure of the model \cite{Kallosh:2022ggf}:
\be
n_s\simeq 1-{2\over N_c}\, \qquad r\simeq{12 \alpha\over N_c^2}\,.
\label{src}\ee

An important point is that $N_c$ is not the physical number of e-folds
elapsed between horizon exit and the end of inflation in the waterfall
model. We denote the latter by $N_*$. The waterfall removes a positive
interval $\Delta N$ of the slow-roll evolution near the end of inflation.
Therefore, for a fixed physical value of $N_*$, the pivot scale leaves the
horizon at a larger field value than in the original model. The same
horizon-exit field value would correspond, in the parent unmodified
$\alpha$-attractor, to
\begin{align}
N_c=N_*+\Delta N .
\label{eq:Nc-definition}
\end{align}
Thus, $N_c$ should be understood as an effective plateau e-fold number,
rather than as the physical duration of inflation. Equivalently, it
measures how far out on the inflationary plateau the pivot scale exits.
The local slow-roll parameters at horizon exit are controlled by this
position, leading to the attractor predictions in
eq.~\eqref{src}. A similar shift of the horizon-exit point induced by a late-time modification of the inflaton dynamics was recently studied using nonminimal derivative coupling in \cite{Fu:2025ciy}. In hybrid $\alpha$-attractor T-models, the corresponding increase of $N_c$ was found in \cite{Kallosh:2022ggf} to be 
\be
N_c\approx N_*+{3\a\over 8} {V_{\rm up} +V_0\over V_0}e^{{\sqrt {2  \over 3 \alpha}} \varphi_c } \ .
\label{NcHyb}\ee
Here $\vp_c$ is the position of premature termination of inflation (for E-models, ${3\a\over 8}$ is replaced by ${3\a\over 4}$) and $V_{\rm up}$ is the value of the uplifted potential in hybrid models. This result, as with the previous equations, has been obtained in the slow-roll approximation in the context of models in which inflation ends when the field $\vp$ reaches the waterfall at $\vp = \vp_{c}$.

Initially, the results of \cite{Kallosh:2022ggf} did not attract much attention since the original $\alpha$-attractors, as well as the Starobinsky model and Higgs inflation, provided a good match to the CMB-related observational data $n_{s} = 0.9682 \pm 0.0032$ \cite{SPT-3G:2025bzu,Balkenhol:2025wms}. However, these models are disfavored at about the 2$\sigma$ level by the CMB-related data combined with the latest DESI DR2 data \cite{DESI:2025zgx,AtacamaCosmologyTelescope:2025blo,SPT-3G:2025bzu}, which yield a higher value of $n_{s}$:\,  $n_{s}=0.9728\pm0.0029$ \cite{SPT-3G:2025bzu,Balkenhol:2025wms}.  

As emphasized in \cite{SPT-3G:2025bzu,Ferreira:2025lrd,McDonough:2025lzo}, this result should be interpreted with caution because DESI DR2 data are in  $\sim 3\sigma$ tension with CMB data. Moreover, the constraints on $n_{s}$ are rather sensitive to the choice of a parameter set.
For example, if one allows variation of some other parameters such as running indices $\alpha_{s}$ and $\beta_{s}$, the tension between the data and the original $\alpha$-attractors,  the Starobinsky model, and Higgs inflation becomes less pronounced, $n_{s}=0.9704\pm0.0038$  \cite{Sabogal:2026qvy}. This tension becomes even smaller if one allows a non-zero $\Omega_{k}$ \cite{Giare:2026oti,Chudaykin:2026amr}, and it entirely disappears if one considers an extended set of parameters including $\sum m_{\nu}$ and $N_{\rm eff}$ \cite{Giare:2026oti}.  On the other hand, some attempts to address the $H_{0}$ tension prefer much larger values of $n_{s}$ \cite{Ye:2022efx,Jiang:2022uyg, Cruz:2022oqk,Giare:2024akf,Giare:2026oti}. Thus it would be nice to have some degree of flexibility in this respect.

Fortunately, hybrid $\alpha$-attractors allow one to describe a broad range of $n_{s}$ without giving up the basic principles of the original $\alpha$-attractor models  \cite{Kallosh:2022ggf,Braglia:2022phb,Kallosh:2025ijd}.  In these models one can reach very high values of $n_{s}$, all the way up to $n_{s} \approx 1$. It is especially clear in the models with $\vp_{c} \ll \sqrt{6\alpha}$, since for $\vp  \ll \sqrt{6\alpha}$ the simplest hybrid $\alpha$-attractor models coincide with the original hybrid inflation models which typically predict $n_{s} \approx 1$  \cite{Linde:1991km,Linde:1993cn}.  In such models one may deviate from the relation $r\simeq 3\a(1-n_s)^2 $, but this is not a problem as long as $r < 0.034$ \cite{Balkenhol:2025wms}, which is very easy to achieve.

\begin{figure}[H]
\vskip 0.5cm 
\centering
		 \includegraphics[width=0.7\textwidth]{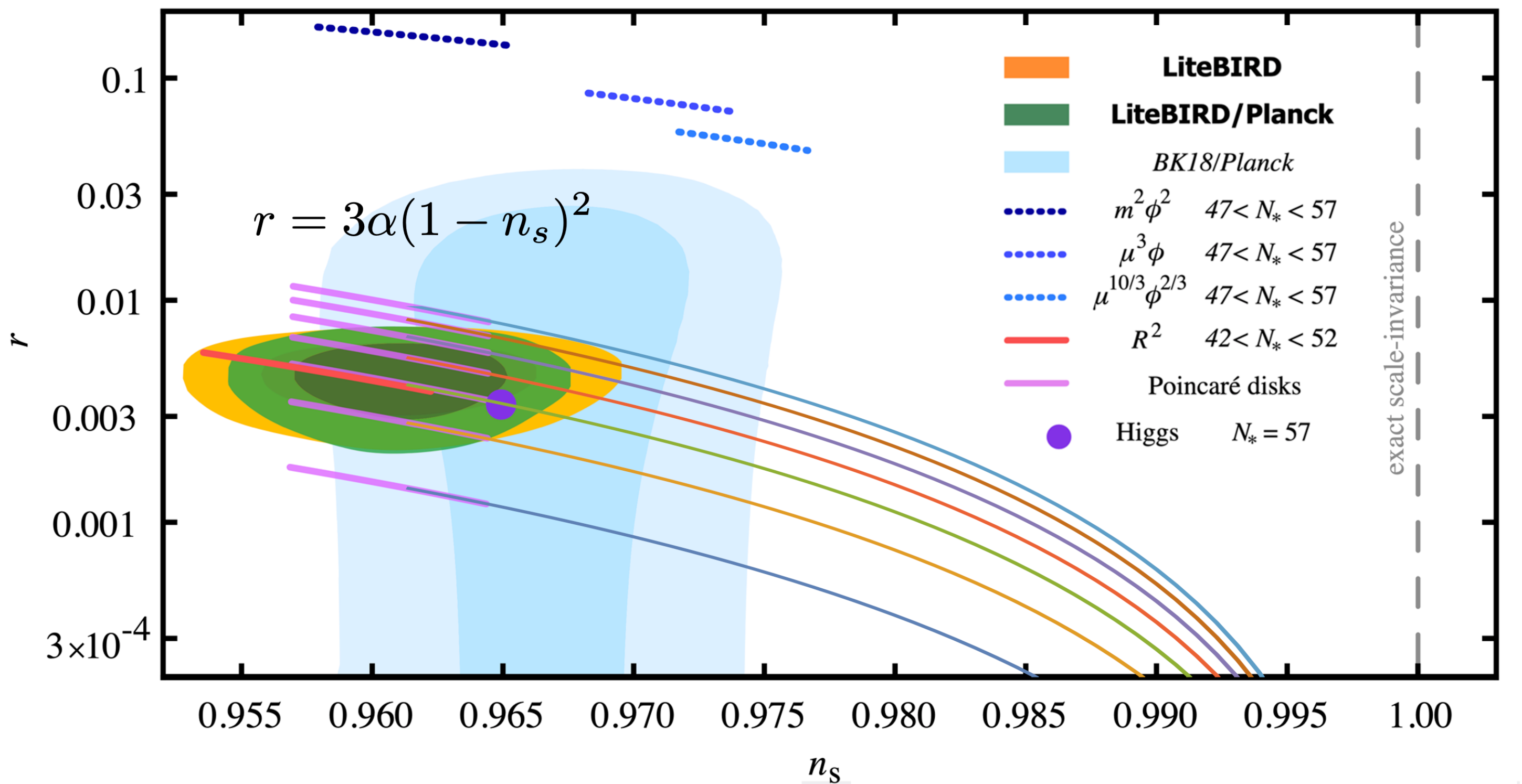}
        \caption{\footnotesize This is an edited version of Fig. 2 of the  LiteBIRD collaboration article  \cite{LiteBIRD:2022cnt}. The seven descending curves show the relation between $r$ and $n_s$
given by ~\eqn{srr}. The purple segments indicate the predictions of the
standard T-model with preferred values of $3\alpha = 1, 2,...,7,$ called ``Poincar\'e disks'' in the LiteBIRD figure. In the versions of hybrid attractors preserving the relation \eqn{srr},  as well as in the single-field $\alpha$-attractor waterfall models introduced
below, the predictions for $n_{s}$ and $r$ move to the right and down along the corresponding curves.}
        \label{Flux2}
\end{figure}

In those models where the relation $r\simeq  3\a(1-n_s)^2 $ is satisfied, the corresponding cosmological predictions are very easy to grasp; see Fig. \ref{Flux2}.
But hybrid inflation is a two-field scenario with several parameters. For each set of parameters, one must check, for example, whether the waterfall transition is sufficiently fast and not followed by yet another stage of inflation, and whether the isocurvature perturbations of the field $\chi$ may affect the final results (which is a very interesting regime that may be responsible for PBH production  \cite{Braglia:2022phb}). Therefore, it would be nice to incorporate a waterfall transition into some single-field models that preserve the advantages of hybrid inflation while being simpler and easier to analyze. This is the goal of our paper.

\section{Waterfalls in single-field inflationary models} \label{Sec:2}

A set of single-field models of the desirable type was recently proposed in \cite{Zhang:2026ivx,Yuan:2026xcg}, and their $\alpha$-attractor generalization was presented in \cite{Chudaykin:2026amr}. In this paper, we develop a large set of single-field $\alpha$-attractors with waterfalls that can describe a continuous spectrum of values of $n_{s}$, all the way up to $n_{s} = 1$. As we will show, some of these models can simultaneously describe inflation and dark energy.

The models introduced in  \cite{Zhang:2026ivx,Yuan:2026xcg} can be represented as follows:
\be
V_{\rm{wf}}(\vp) = { V_{\rm original}(\vp) \over 1+\gamma} \Big   ( 1+\gamma \tanh {\varphi-\varphi_c\over \Delta \varphi}\Big) \ .
\label{cases}
\ee

Here $V_{\rm original}(\vp)$ is some inflationary potential, modulated by the function $1+\gamma \tanh {\varphi-\varphi_c\over \Delta \varphi}$. For convenience, we introduced here the coefficient $1\over 1+\gamma$, so that at $\varphi-\varphi_c \gg \Delta \vp$ the potential $V_{\rm{wf}}(\vp)$  practically coincides with $V_{\rm original}(\vp)$. Meanwhile, at $\varphi_{c}-\varphi \gg \Delta \vp$ the potential becomes smaller than $V_{\rm original}(\vp)$ by ${2\gamma\over 1+\gamma} V_{\rm original}(\vp)$. Most of the change occurs in the vicinity of $\vp = \vp_{c}$ of width $\Delta \vp$.  In the limit $ \Delta \varphi \ll \varphi_c$, this modification looks like an instant step down, a waterfall of height  ${2\gamma\over 1+\gamma} V_{\rm original}(\vp_{c})$. 

The investigations in \cite{Zhang:2026ivx,Yuan:2026xcg} used the following range of parameters
\be
  \gamma \leq  1 , \quad  0.04 \leq \Delta \vp \leq  0.286\,.
\label{inst}\ee
The case with a moderate waterfall-type step down with $\gamma=0.1, \, \Delta \vp= 0.286$ was studied in~\cite{Zhang:2026ivx} and allowed $n_s$ to reach $n_{s }\approx 0.975$ in some models. 
The case of a step with an instant waterfall with $\gamma=1, \Delta\varphi =   0.04$ was studied in \cite{Yuan:2026xcg}, and allowed one to reach much higher $n_s$, up to $n_{s }\approx 0.996$. 

The potential $V=V_{\rm original} \cdot V_{\rm step}$, where $V_{\rm step}= 1+\gamma \tanh {\varphi-\varphi_c\over \Delta \varphi} $, was introduced long ago in \cite{Adams:2001vc} and was often used in cosmological applications.  However, it was not clear whether one could implement such functions in $\alpha$-attractors. One way to do it will be described in the next section.   

In the subsequent sections, we will also introduce a new set of models constructed in a similar but different way:  $V=V_{\rm original} + \gamma V_0 \left (1+\tanh {\varphi-\varphi_c\over \Delta \varphi}\right)$.

\section{Waterfall $\a$-attractors } \label{Sec:3}
\subsection{Hyperbolic geometry}

The original E- and T-models in hyperbolic geometry can be represented in a form in which the potentials depend on a half-plane variable $T$ \cite{Kallosh:2026tjm}. They have the kinetic term 
\be
-3\a {\partial T \partial\bar T\over (T+\bar T)^2}\,,
\ee
and potentials
\be
V^{E}(T, \overline{T}) =V_0  [ (1-T) (1-\overline{T})]^{n} \ ,
\label{VEE}\ee
\be
V^{T} (T, \overline{T})= V_0 \left [ {1-T\over 1+T} \, { 1-\overline{T}\over 1+\overline{T}} \right ]^{n} \ .
\label{VET}\ee
The geometric half-plane complex coordinate  $T$ for  an inflaton-axion pair is
\be 
T=e^{-{\sqrt {2  \over 3 \alpha}} \varphi }+i\theta\, , \qquad     {\vp\over \sqrt 6\a}  =-{1\over 2} \ln {T+\overline{T}\over 2} \ .
\ee
The waterfall-modulated potentials in \eqn{exampleT} and  \eqn{exampleE} now are

\be\label{tnu}
V= {V^{T, E}\over 1+\gamma}  \cdot  \left(1+ \gamma\  { {1-({T+\overline{T}\over 2T_c})^\nu}\over {1+({T+\overline{T}\over 2T_c})^\nu }
}\right),
\ee
where we take $\gamma>0$ and
\be\label{nu}
T_c\equiv e^{{-\sqrt {2  \over 3 \alpha}} \varphi_c }\, , \quad \nu = {\sqrt{6\a}\over \Delta \vp}.
\ee
Using this relation, one can show that
\be
V= {V^{T, E}\over 1+\gamma}   \cdot  \Big[1+ \gamma\ \tanh  {\varphi-\varphi_c\over \Delta \varphi}\Big]  \ .
\ee
Alternatively, using \rf{nu}, we find
\be\label{nuup}
V= {V^{T, E}\over 1+\gamma}  \cdot  \left[1+ \gamma\ \tanh { \nu(\varphi-\varphi_c)\over  \sqrt{6\alpha}}\right] \ .
\ee
For the simplest T- and E-models with $n = 1$, we find waterfall potentials
\be
V_{\rm {wf}}^T= {V_0\over 1+\gamma}\, \left (\tanh {\vp\over  \sqrt{6\a} }\right)^{2} \, \left( 1+\gamma \tanh {\varphi-\varphi_c\over \Delta \varphi}\right) \ ,
\label{exampleT}\ee
\be
V_{\rm{wf}}^E= {V_0\over 1+\gamma}\, \left (1- e^{-\sqrt{2\over 3\a} \vp}\right)^2 \, \left ( 1+\gamma \tanh {\varphi-\varphi_c\over \Delta \varphi}\right) \ .
\label{exampleE}\ee
The increase of the potentials due to the added term $\gamma \tanh {\varphi-\varphi_c\over \Delta \varphi} $ is compensated here by a factor ${1\over 1+\gamma}$ so that at $\vp\to \infty$ all models have the same plateau.

We will find here that there is a choice of the waterfall position $\vp_c$ that allows one {\it to move along the descending horizontal curves in Fig. \ref{Flux2} continuously} from values of $(n_s, r)$ in original $\a$-attractor models up to very large values of $n_s$ and smaller $r$.  In our examples, we vary the waterfall from a very sharp, high step, with $\Delta\varphi=0.04$ and $\gamma=1$, to broader, lower steps with $\Delta\varphi$ up to $0.25$ and smaller values of $\gamma$.

We will present these results in the context of hyperbolic geometry. The two-shoulder or two-plateau potentials are very natural in hyperbolic geometry, where the examples were given in \cite{Carrasco:2015uma} in the context of initial conditions for inflation, and in quintessential $\a$-attractors describing inflation and evolving dark energy in \cite{Dimopoulos:2017zvq,Akrami:2017cir,Zhumabek:2023wka,Jing:2026ymp}. Here, again, we insert a two-shoulder potential into a simple $\a$-attractor potential, starting outside of the inflationary plateau and moving gradually deeper into the inflationary plateau. This results in a continuous increase of  $n_s$, a decrease of  $r$, as shown in Fig. \ref{Flux2}.


\subsection{Evaluation of $N_c$}\label{eval}
The detailed evaluation of the effective 
\be
N_c= N_*+ \Delta N
\ee in waterfall $\a$-attractor T- and E-models is performed in the Appendix 
\ref{App:eval}. The expression for $\Delta N$ is always positive.

 We distinguish two regimes. In the strong regime, inflation terminates while the waterfall contribution still dominates the slope. In the weak regime, the inflaton passes through the waterfall-dominated region before inflation ends. For the $n=1$ T-model with a strong waterfall located sufficiently high on the plateau, we find 
\begin{align}
N_c^{\rm strong} \simeq N_* + {3\a\over 8}  \, e^{{\sqrt {2  \over 3 \alpha}} \varphi_c } \, A_{\rm wf}^{\rm strong} \,,
\label{st1}\end{align} 
\be
A_{\rm wf}^{\rm strong} = Q_+^{\frac{\kappa}{2-\kappa}}\,,
\label{st2}\ee
\begin{align}
 Q_\pm\equiv\frac{\gamma   \sqrt{\frac{3\alpha}{2}}  \, e^{{\sqrt {2  \over 3 \alpha}} \varphi_c }}{(1 \pm\gamma)\Delta\varphi\, } , \qquad {\frac{\kappa}{2-\kappa}}= { \sqrt{\frac{2}{3\alpha}} \Delta\varphi\over 2-\sqrt{\frac{2}{3\alpha}} \Delta\varphi}\,.
\label{st3}\end{align}
If the inflaton passes through the waterfall-dominated region before inflation ends, we find
\begin{align}
N_c^{\rm weak} \simeq N_* + {3\a\over 8}  \, e^{{\sqrt {2  \over 3 \alpha}} \varphi_c } \, A_{\rm wf}^{\rm weak} \,,
\label{we1}\end{align}
\begin{align}
A_{\rm wf}^{\rm weak} = \left[Q_+^{\frac{\kappa}{2-\kappa}}-Q_-^{-\frac{\kappa}{2+\kappa}}\right].
\label{we2}\end{align}
In a two-field hybrid $\a$-attractor \cite{Kallosh:2022ggf}, the relevant formula can be given in the form
\be
N_c\approx N_*+{3\a\over 8} \,  e^{{\sqrt {2  \over 3 \alpha}} \varphi_c }\, A^{\rm hybrid}\,,
\label{h1}\ee
\be
A^{\rm hybrid} = {V_{\rm up} +V_0\over V_0}\,.
\label{h2}\ee
In all cases the factor ${3\a\over 8}  \, e^{{\sqrt {2  \over 3 \alpha}} \varphi_c } $ is the same; it depends on $\a$ and on the transition point $\vp_c$.  However, in hybrid $\alpha$-attractor models  \cite{Kallosh:2022ggf}  $A^{\rm hybrid} $ depends on $V_{\rm up}$ defined by the properties of the waterfall field, responsible for the premature termination of inflation. In single-field models, which we study here, both factors $A_{\rm wf}^{\rm strong}$ and $A_{\rm wf}^{\rm weak}$ depend on the waterfall's features and position.

The effective number of e-folds $N_c$ entering the attractor expressions \eqn{src} is therefore higher than that in models without waterfalls, $N_c> N_*$. Therefore, $n_s$ increases and $r$ decreases due to the waterfalls, as we show in Fig. \ref{Flux2}.

\section{Moving  along the descending curves in Fig. \ref{Flux2}}\label{Sec:4}
\subsection{Waterfall-modulated T- and E-models, in examples}

\subsubsection{T-models, instantaneous waterfall, $\alpha = 1$, $\gamma=1$, $\Delta\vp=0.04$}\label{sec:examplees}
\begin{figure}[H]
\vskip 0.5cm 
\centering
		 \includegraphics[width=0.54\textwidth]{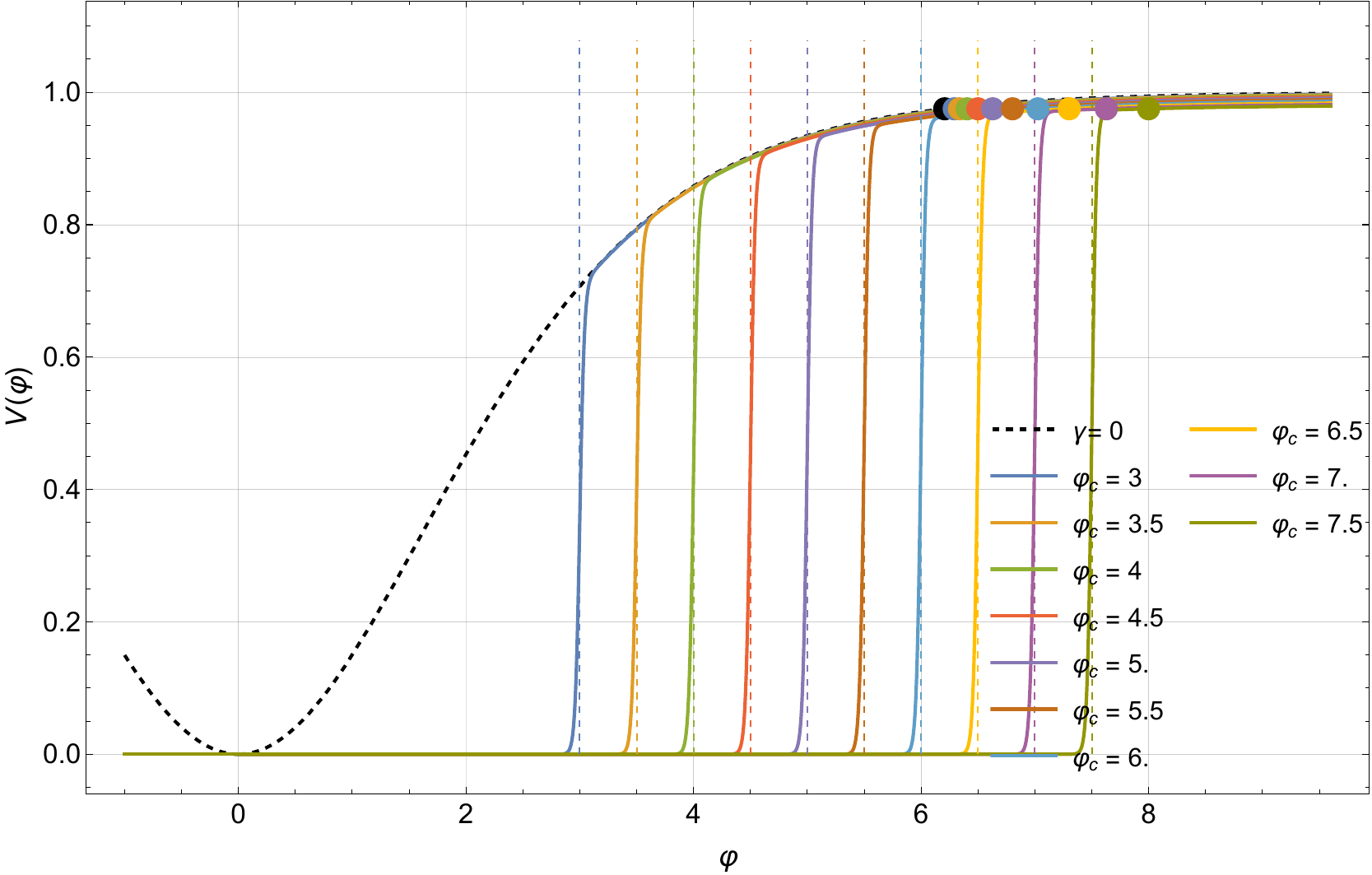}
        \caption{\footnotesize The black dashed line is a $\gamma=0$ T-model $\a=1$ potential without a waterfall. We also show the shape of the T-model potential for the instantaneous waterfall case with $\gamma=1$, $\Delta\varphi=0.04$. Each marker corresponds to the value of $\varphi_*$ for $N_*=60$. The potential with any $\vp_c$ instantly drops to zero.}
        \label{PotT}
\end{figure}

\begin{figure}[H]
\vskip 0.3cm 
\centering
		 \includegraphics[width=0.57\textwidth]{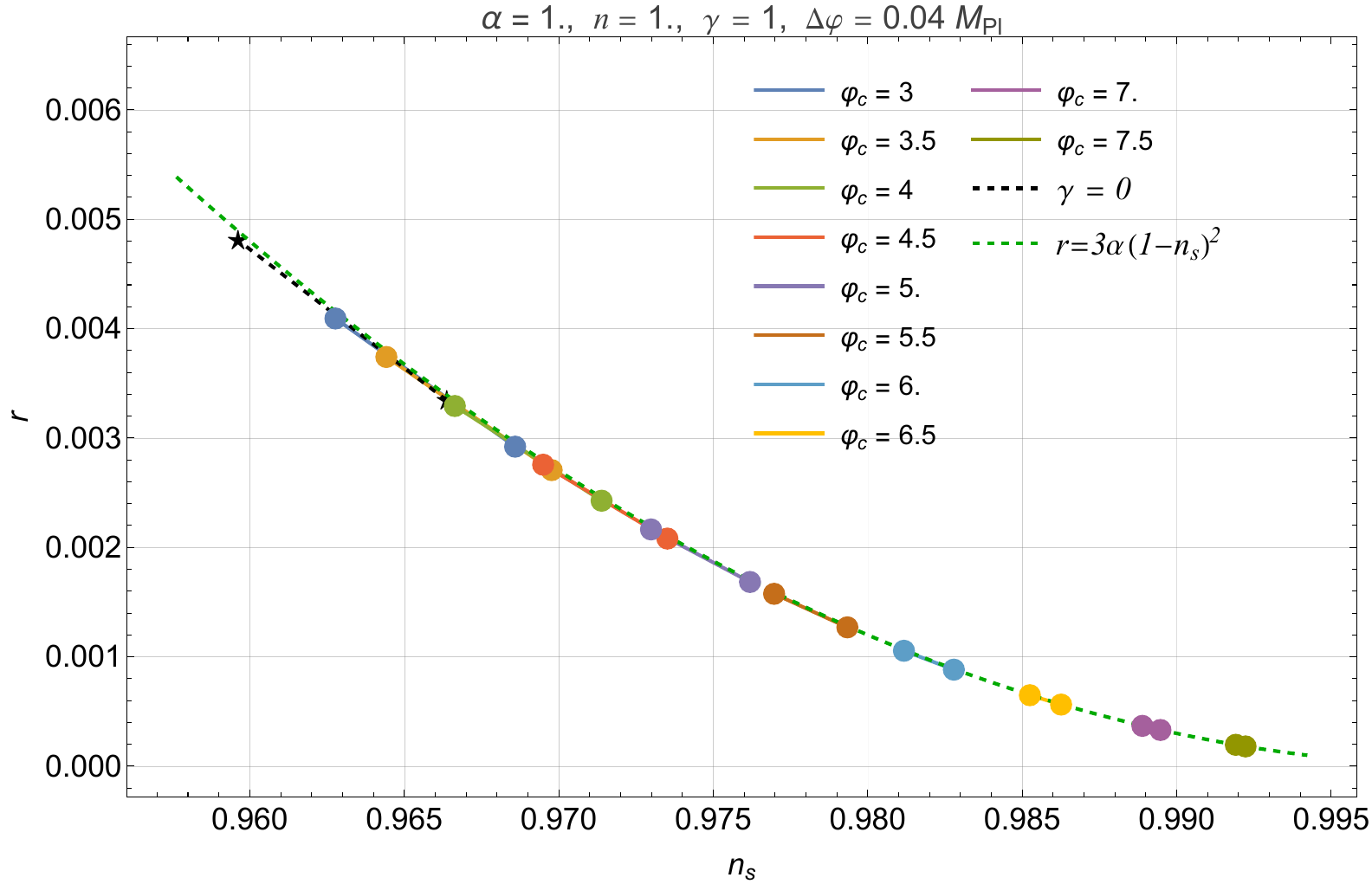}
        \caption{\footnotesize         The black dashed curve shows the predictions of the $\alpha=1$ T-model with $\gamma=0$, corresponding to the model without a waterfall. 
        The green dashed line is defined by Eq. $r=3(1-n_s)^2$. Two markers of the same color correspond to $N_*=50, 60$. Thus, with more choices of  $\vp_c$, we can densely populate the total green curve and reach $n_s\approx 0.992$.}
        \label{nsr}
\end{figure}

To verify our theoretical expectations, we evaluated $n_{s}$ and $r$ for the T-model \rf{exampleT}. The corresponding potentials and the $(n_s,r)$ predictions are shown in Fig. \ref{PotT} and Fig. \ref{nsr}.

As one can see, the results for T-models nearly perfectly match the theoretical expectation $r=3(1-n_s)^2$.
The range of waterfall positions for which the predictions remain close to the curve $r=3\alpha(1-n_s)^2$ is limited. If the waterfall is placed too deep in the inflationary region, inflation terminates too early for a given choice of the waterfall parameters, and the resulting $(n_s,r)$ values begin to deviate from the attractor curve.

\subsubsection{Wider waterfall, $\alpha = 1$, $\gamma=1$, $\Delta\vp=0.25$}\label{sec:wide}

\begin{figure}[H]
\centering
	\includegraphics[width=0.56\textwidth]{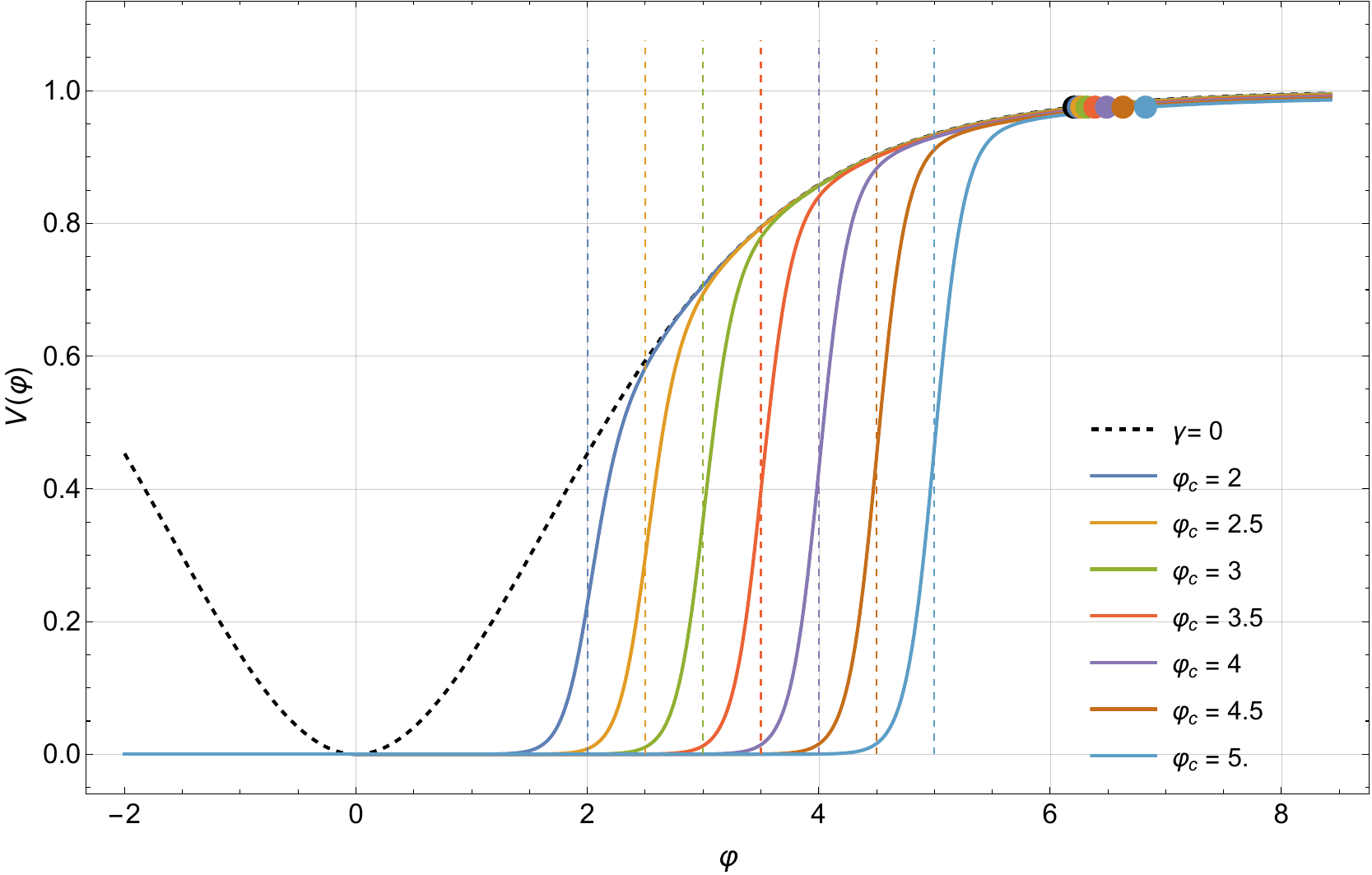}
        \caption{\footnotesize The potential \rf{exampleT}  for wider waterfalls, $\Delta\vp  = 0.25$ instead of $0.04$}
        \label{PotTw}
\end{figure}

\begin{figure}[H]
\centering
		 \includegraphics[width=0.58\textwidth]{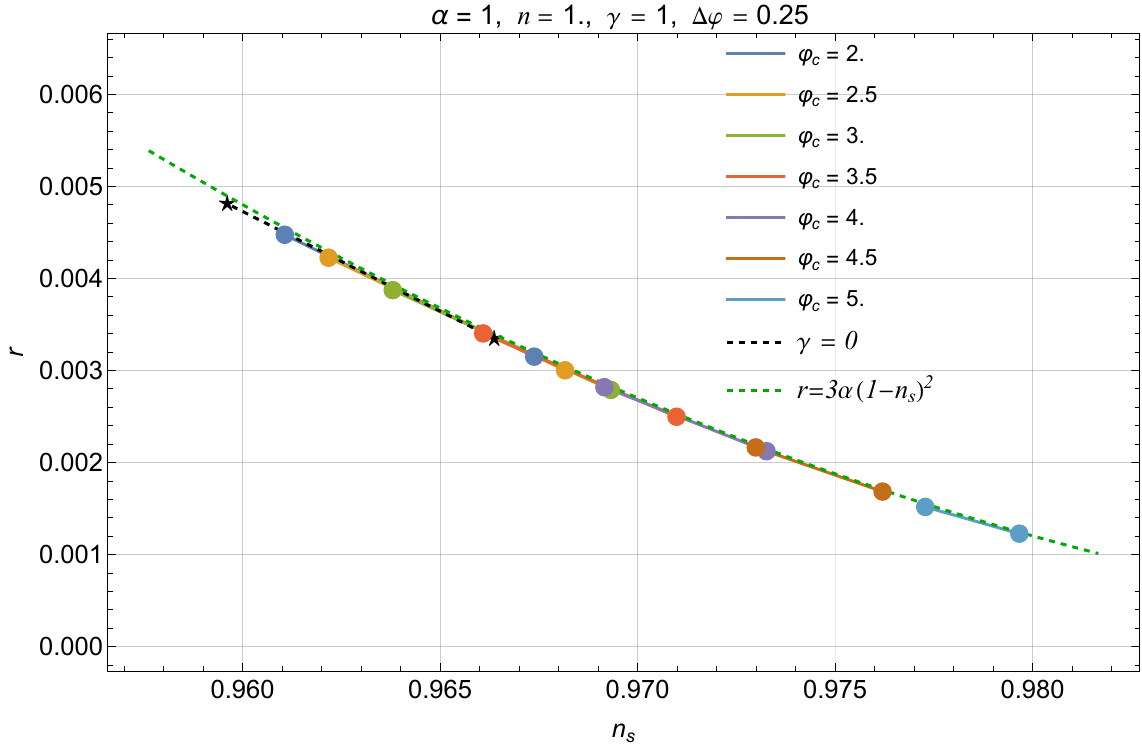}
        \caption{\footnotesize The difference from Fig. \ref{nsr} is that $\Delta\vp=0.04$ is now replaced by $\Delta\vp=0.25$ and the positions $\vp_c$ have changed. One can reach $n_s\approx 0.98$.}
        \label{nsrw}
\end{figure}

Here we will show the results for the wider waterfalls, $\Delta\vp  = 0.25$ instead of $0.04$. The corresponding potentials and the $(n_s,r)$ predictions are shown in Fig. \ref{PotTw} and Fig. \ref{nsrw}.  

\subsubsection{Smaller height  waterfall, $\gamma=0.3$, $\Delta\vp=0.25$}

\begin{figure}[H]
\vskip 0.3cm 
\centering
		 \includegraphics[width=0.55\textwidth]{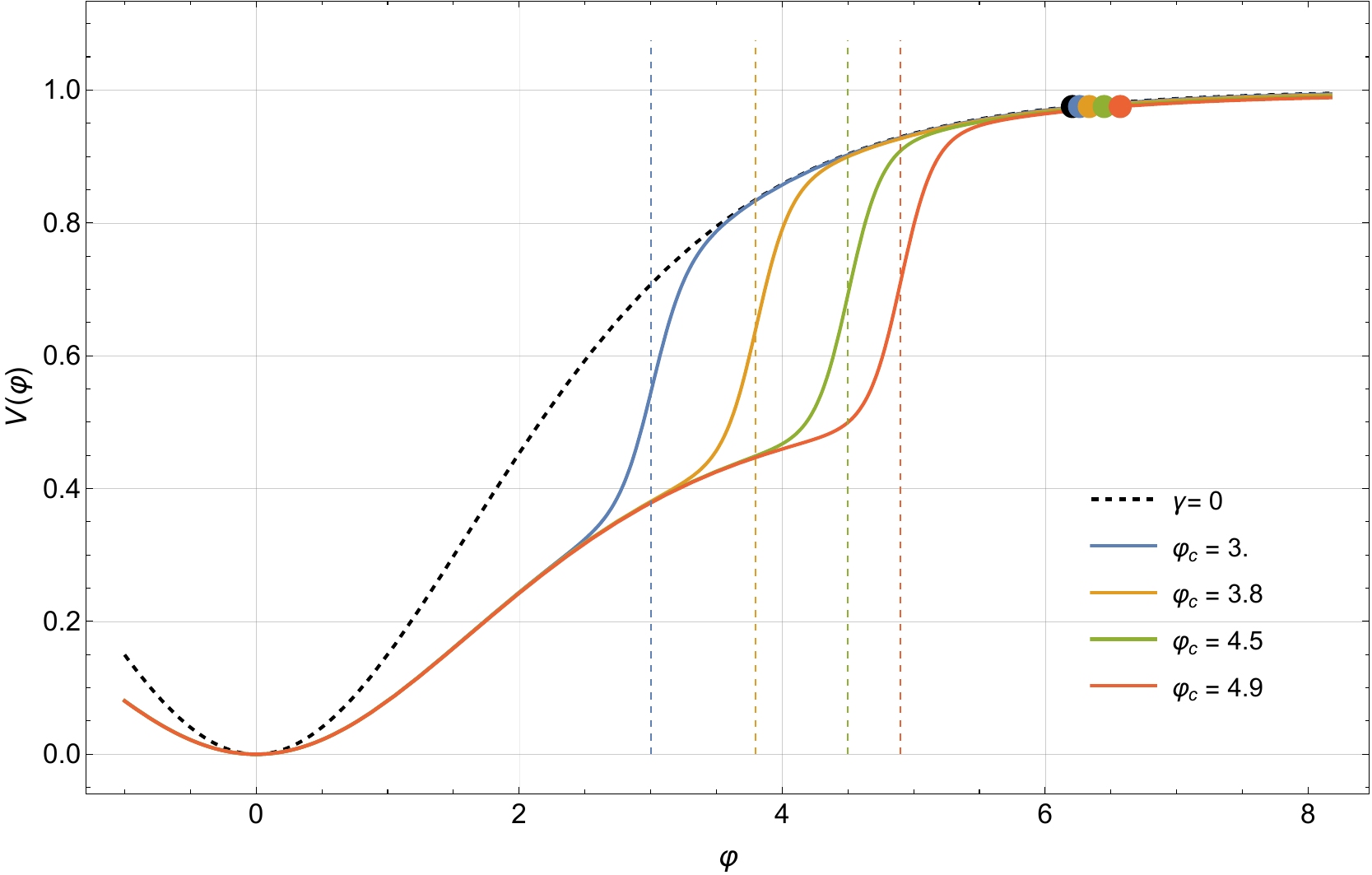}
        \caption{\footnotesize The difference from Fig. \ref{PotTw} is that the height is smaller, $\gamma=0.3$, and the positions $\vp_c$ have changed.}
        \label{Pot}
\end{figure}

\begin{figure}[H]
\vskip 0.3cm 
\centering
		 \includegraphics[width=0.6\textwidth]{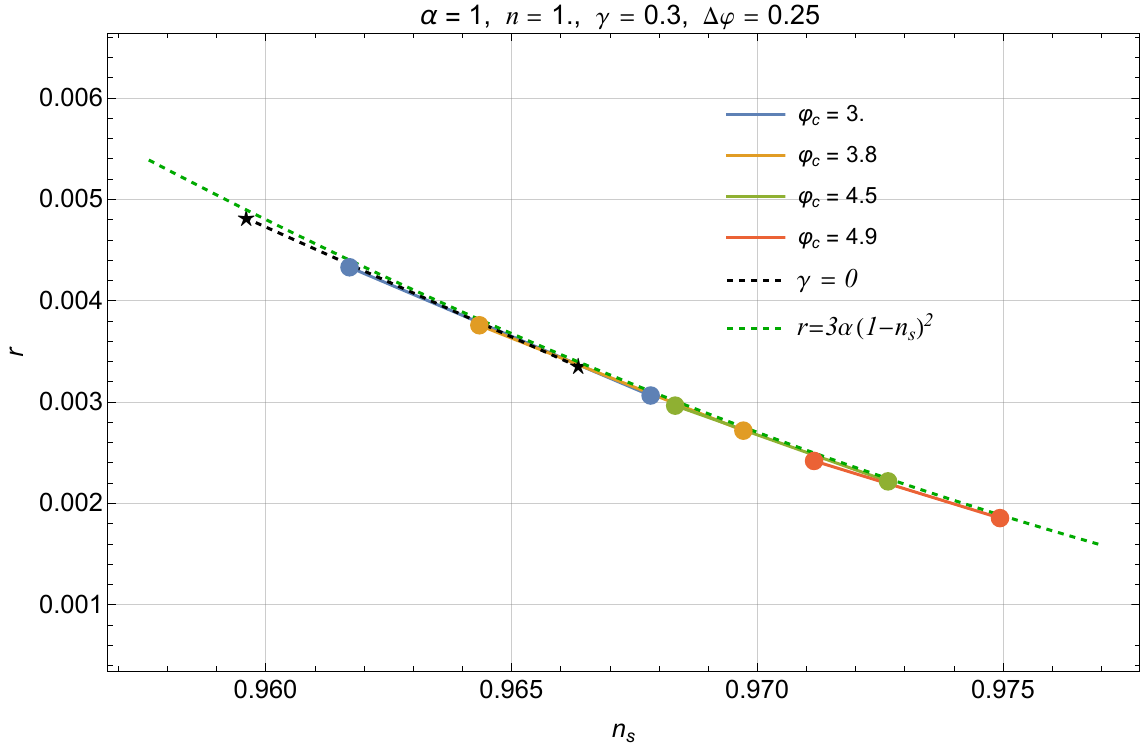}
        \caption{\footnotesize The difference from Fig. \ref{nsrw} is that now $\gamma=0.3$ and the positions $\vp_c$ have changed. In this case one can reach up to $n_s\approx 0.975$. }
        \label{nsrInter}
\end{figure}

The results are presented in Figs. \ref{Pot} and \ref{nsrInter}.

\subsubsection{E-model, instantaneous waterfall, $\alpha = 1$, $\gamma=1$, $\Delta\vp=0.04$}
\begin{figure}[H]
\vskip 0.3cm 
\centering
		 \includegraphics[width=0.6\textwidth]{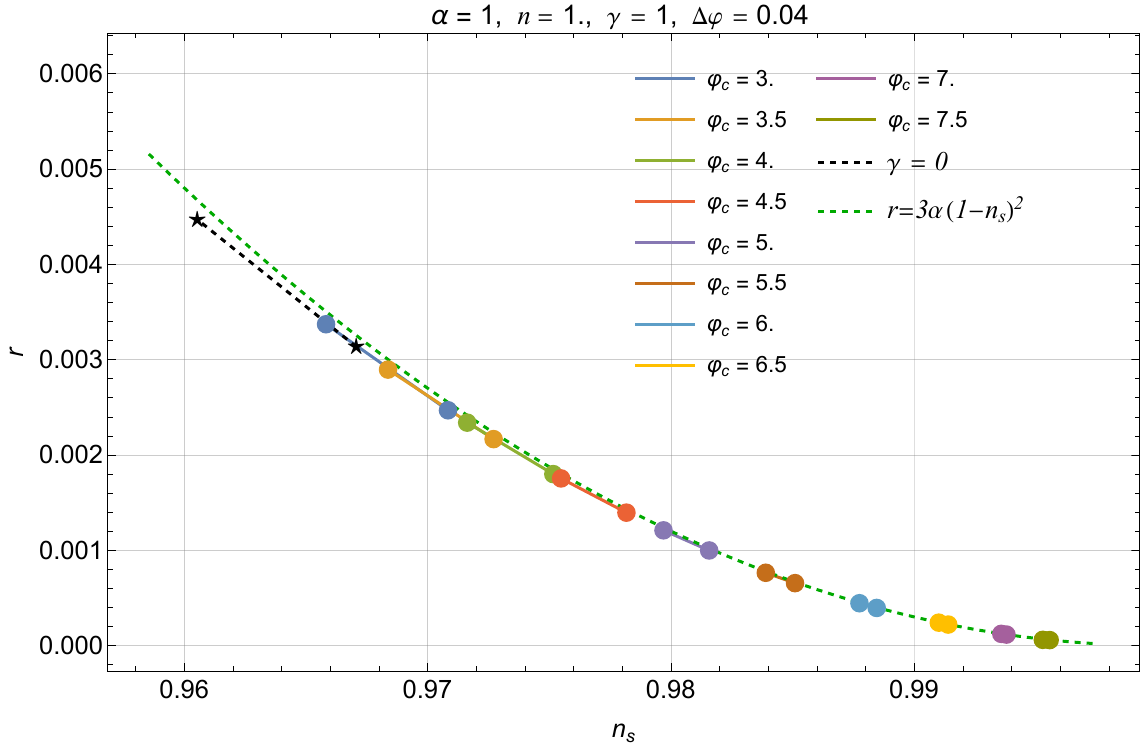}
        \caption{\footnotesize        The black dashed curve shows the predictions of the $\alpha=1$ E-model
with $\gamma=0$, corresponding to the model without a waterfall. The green dashed line is defined by eq. $r=3(1-n_s)^2$. Two markers of the same color correspond to $N_*=50, 60$. With more choices of intermediate $\vp_c$, one can densely populate the entire green curve. However, at small $n_{s}$, the results deviate slightly from this curve.  }
        \label{nsrE}
\end{figure}

\subsubsection{ Wider waterfall, $\gamma=1$, $\Delta\vp=0.25$}

\begin{figure}[H]
\vskip 0.3cm 
\centering
		 \includegraphics[width=0.6\textwidth]{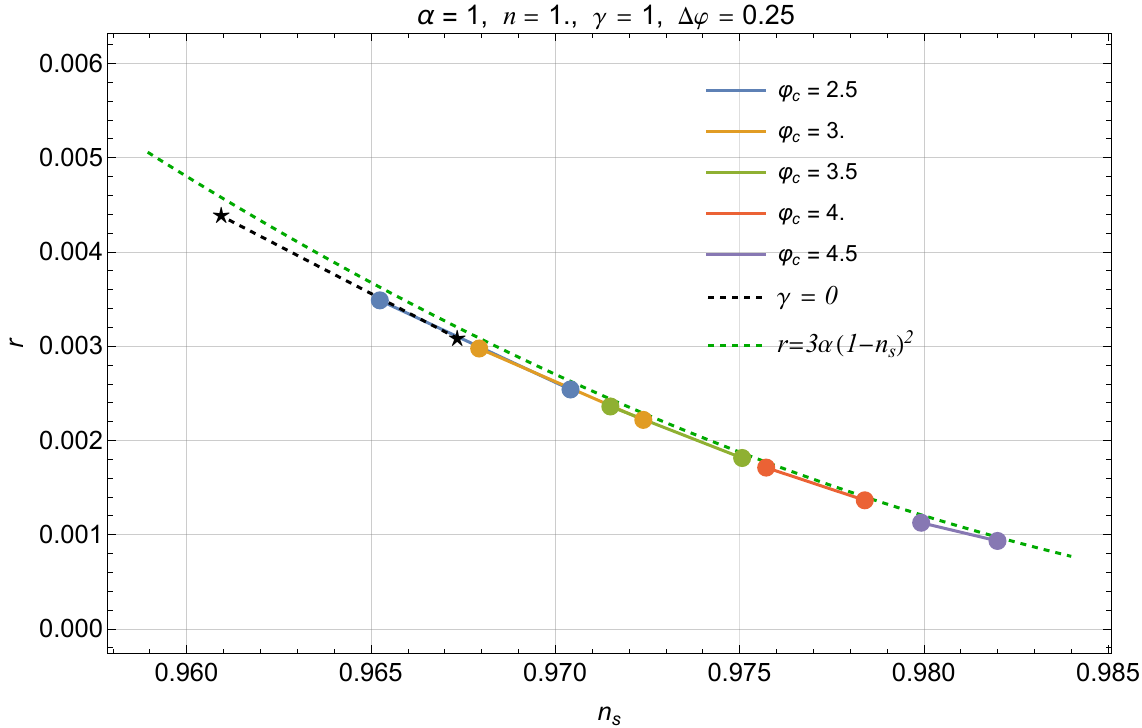}
        \caption{\footnotesize Same as Fig. \ref{nsrE}, but $\Delta \vp=0.04$ is replaced by $\Delta \vp=0.25$ and $\vp_c$ has changed.}
        \label{nsrA}
\end{figure}

\subsubsection{Wider, smaller height waterfall}

\begin{figure}[H]
\vskip 0.3cm 
\centering
		 \includegraphics[width=0.65\textwidth]{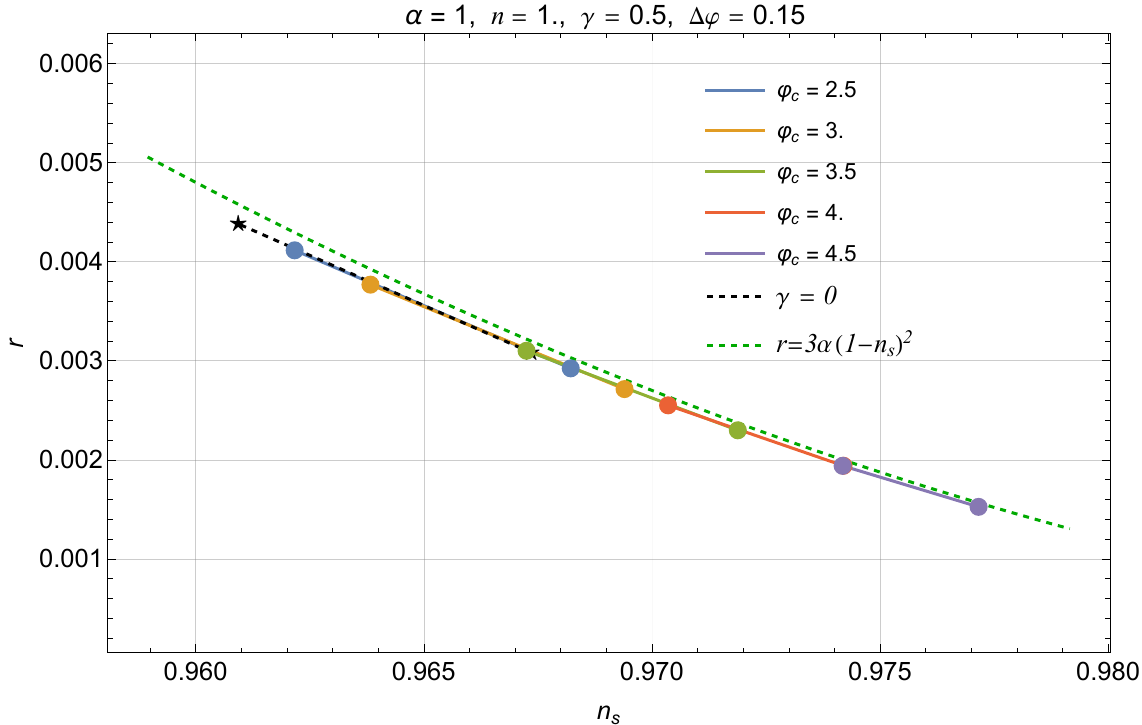}
        \caption{\footnotesize Here we take a wider waterfall with $\Delta \vp=0.15$ and a smaller height with $\gamma=0.5$.}
        \label{nsrA}
\end{figure}
\noindent Thus we have shown various choices of parameters in waterfall-modulated $\a$-attractors that provide examples
corresponding to Fig. \ref{Flux2}, with these models starting with the $\a=1$ Poincar\'e disk and moving toward increasing $n_s$.

\subsection{Comparing the waterfall evaluation  with  numerical examples}
 We now compare the numerical results with the analytic estimate of
$\Delta N$ shown in Sec.~\ref{eval}. We denote by $n_s^{\rm app}$ the value
obtained from $n_s=1-2/N_c$ using the analytic estimate of $N_c$, and by
$n_s^{\rm num}$ the corresponding numerical result, which is derived by numerically solving the full equation of motion.

\fbox{$\gamma=1,\Delta\varphi=0.04,\varphi_c=7.5,\alpha=1$ T-model}: In this case, inflation terminates while the waterfall contribution still dominates the slope, so the model belongs to the strong regime described by eq.~\eqref{strong N2}, which gives
\begin{align}
\Delta N\approx 198.314 \Rightarrow \ n_s^{\rm app}=0.992257, \quad n_s^{\rm num}=0.992227\,.
\end{align}

\fbox{$\gamma=1,\Delta\varphi=0.25,\varphi_c=5,\alpha=1$ T-model}: As in the previous case, this model belongs to the strong regime.
\begin{align}
\Delta N\approx 39.1542 \Rightarrow \ n_s^{\rm app}=0.9798, \quad n_s^{\rm num}=0.9797\,.
\end{align}

\fbox{$\gamma=0.3,\Delta\varphi=0.25,\varphi_c=4.9,\alpha=1$ T-model}: This case belongs to the weak regime described by \eqref{weak N}. Then, we find
\begin{align}
\Delta N\approx 19.5356 \Rightarrow \ n_s^{\rm app}=0.9749, \quad n_s^{\rm num}=0.9749\,.
\end{align}
These results show the validity of our estimate of $\Delta N$.

Additional examples shown in Figs.~\ref{nsr}, \ref{nsrw}, and \ref{nsrInter} demonstrate that,
for $\alpha=1$, varying the waterfall position $\phi_c$ densely populates
the curve $r\simeq3\alpha(1-n_s)^2$, although the predictions deviate slightly from it at smaller $n_s$. 
Thus we have shown models with various choices of parameters in waterfall-modulated $\a$-attractors which give predictions  
matching the predictions shown in Fig. \ref{Flux2}  for the particular case $\alpha = 1$.

\section{Uplift and waterfall }\label{Sec:5}

 As discussed in Sec.~\ref{Sec:1}, hybrid $\alpha$-attractors have two effects that can increase $n_s$: an uplift of the inflationary potential and a premature termination of inflation. In Secs.~\ref{Sec:2}--\ref{Sec:4} we concentrated on the latter effect and constructed single-field models in which a multiplicative waterfall modulation modifies the evolution near the end of inflation.

Here we explore a complementary possibility and construct single-field realizations of the uplift mechanism. We consider potentials of the form $V=V_{\rm original}+V_{\rm uplift}$. The basic idea, explained around eq.~\eqref{nsslow}, is that an approximately constant positive contribution raises $V$ without significantly changing $V'$ and $V''$, thereby reducing the relative slope and curvature of the potential. This tends to increase $n_s$ and decrease $r$.

The uplifting terms considered below are not exactly constant functions of $\varphi$. Near the transition region, they also contribute to $V'$ and $V''$. However, on the inflationary side of the transition they approach an approximately constant positive contribution and reproduce the essential uplift effect of hybrid $\alpha$-attractors. In this sense, the models studied in this section provide a single-field realization of the uplift mechanism, complementary to the premature-end mechanism considered in the previous sections.

\begin{figure}[H]
\vskip  5pt
\centering
		 \includegraphics[width=0.5\textwidth]{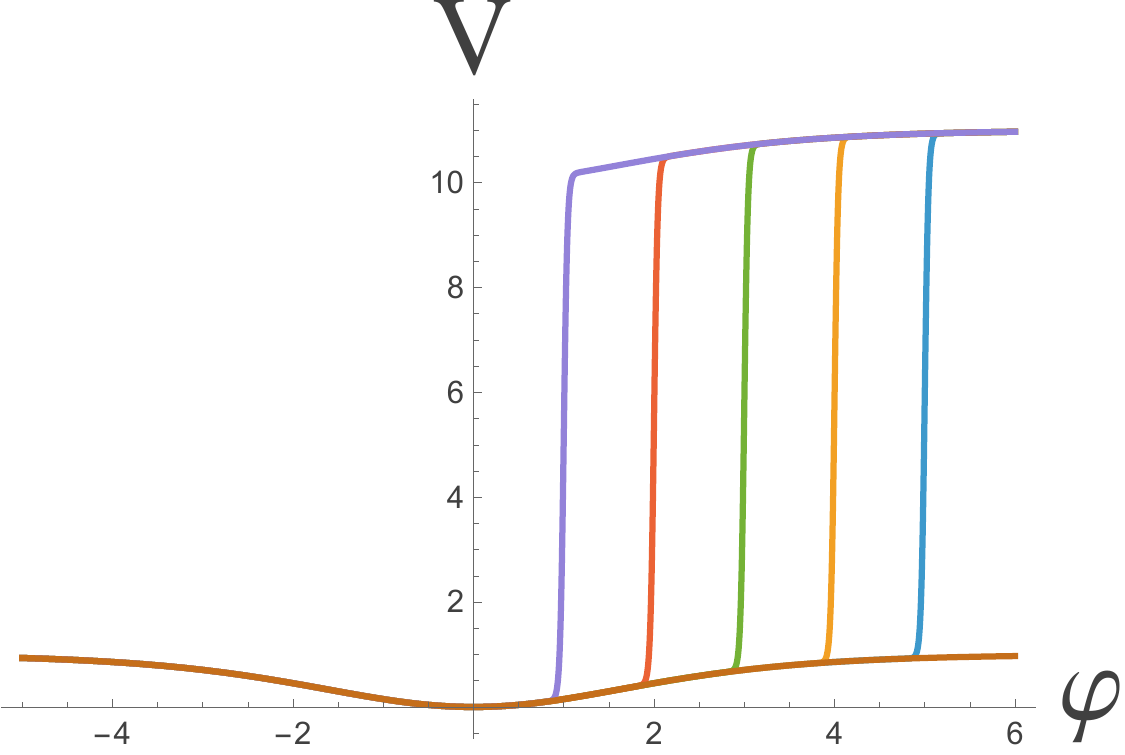}
        \caption{\footnotesize The potential \rf{Tup} for $V_{0} = 1$, $\alpha = 1$, $\gamma = 5$, $\Delta  \vp = 0.04$,  and waterfalls at $\vp_{c}=  1,2,3,4,5$. The lower line shows the original T-model potential. The uplifted potentials join into a single line $\tanh^2{\vp\over\sqrt{ 6\a}} +10$ at large $\vp$.}
        \label{uppot}
\end{figure}

 As a first example, consider a set of models
\be
V_{\rm {wf}}^T= {V_0}\left(\tanh^2{\vp\over\sqrt{ 6\a}} + {\gamma}   \Big  (1+\tanh  {\varphi-\varphi_c\over \Delta \varphi} \Big)\right) \ .
\label{Tup}\ee
The first term is the standard T-model potential. The second term describes the uplift. In the large $\vp$ limit, it uplifts the original potential by $2V_{0} \, \gamma$. At large negative $\vp$, the uplifting term disappears. We show the modified potential in Fig. \ref{uppot} for $V_{0} = 1$, $\gamma = 5$, $\Delta  \vp = 0.04$,  and $\vp_{c}=  1,2,3,4,5$. Note that the depth of the minimum in this particular model is slightly modified because of the uplifting term. Therefore, one should subtract an exponentially small constant from the model to compensate for this effect.

A numerical investigation shows that the waterfall potential with $\vp_c = 1$, $N_* = 55$, $\gamma = 20$ leads to $n_{s} = 1.000$, $r = 0.0004$. For $\gamma = 10$ one has $n_{s} = 0.9931$, $r = 0.0017$.  For $\gamma = 5$ one has  $n_{s} = 0.9779$, $r = 0.0040$.   For $\gamma = 2.5$ one has  $n_{s} = 0.9658$, $r = 0.005$.

These sets of numbers slightly deviate from the attractor relation $r\simeq  3\a(1-n_s)^2$. For example, this condition would imply that $r = 0$ for $n_{s }= 1$, whereas we found $r = 0.0004$. The reason is that $\vp_{c} = 1 $ is far from the $\alpha$-attractor plateau, which, for $\alpha = 1$,  begins at $\vp > \sqrt{6}$. The second reason is that the waterfall for $\gamma = 20$ is very steep, so the slow-roll conditions do not work here.   When the waterfall is located at $\varphi_c\gg\sqrt{6}$  and smaller values of $\gamma$, the  attractor relation   $r\simeq 3\a(1-n_s)^2$ becomes valid.

In particular, for $\vp_{c} = 5$, we found that one has $n_{s} = 0.995$, $r = 0.0001$ for $\gamma = 5$, and the relation $r\simeq 3\a(1-n_s)^2$ is approximately satisfied for $\gamma < 2$. For $\gamma = 2$ we have $n_{s} = 0.9895$, and  $r = 0.0003$.

\

\begin{figure}[H]
\vskip  5pt
\centering
		 \includegraphics[width=0.5\textwidth]{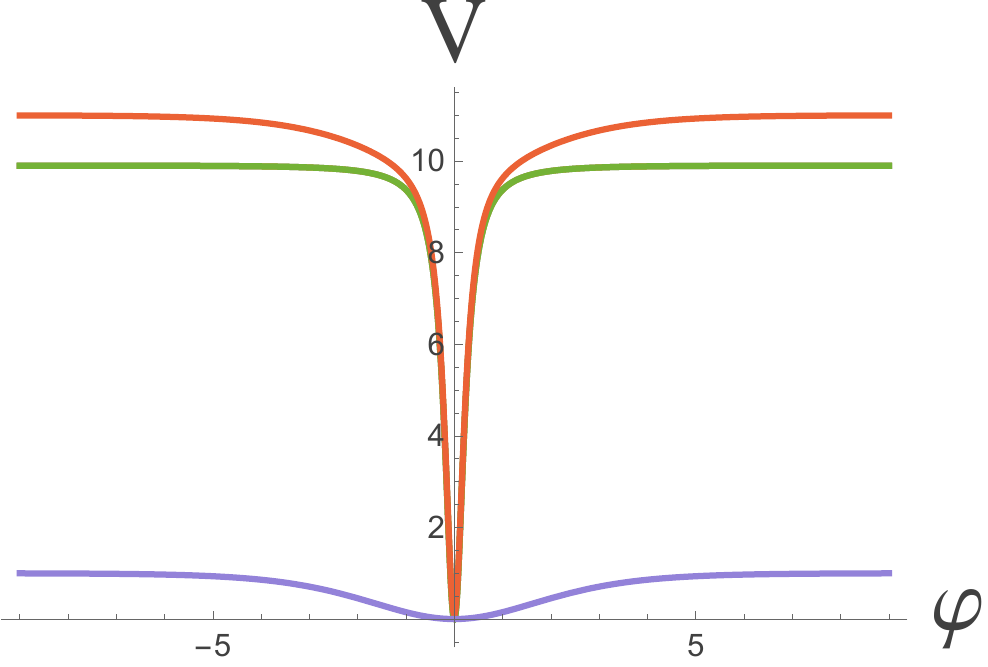}
        \caption{\footnotesize The red line shows the  potential \rf{polstep} for $V_{0} = 1$, $\alpha = 1$, $\gamma = 10$.  The lower dark blue line shows the original T-model potential. The green line shows the uplifting potential, which is the second term in \rf{polstep}.}
        \label{A10}
\end{figure}

The second class of models is based on an uplifting term represented by polyattractor potentials introduced in \cite{Kallosh:2026tke}.  The preceding example combines an approximately constant uplift on the inflationary plateau with a relatively sharp transition. To show that the uplift mechanism does not rely on such a sharp step, we next consider a smoother realization based on polyattractor potentials: 
\be\label{polstep}
V= V_{0} \left( \tanh^{2}{\vp\over \sqrt{6\alpha}} +  \gamma\,   {\tanh^{2}{\vp\over \sqrt{6\alpha}}\over \tanh^{2}{\vp\over \sqrt{6\alpha}}  +d^{2}} \right) \ .
\ee
We show this potential in Fig. \ref{A10} for $V_{0} = 1$, $\alpha = 1$, $\gamma = 10$.

As an example, we take $d = 0.1$, and evaluate $n_{s}$ and $r$ for various values of $\gamma$. We find  $n_{s}=   0.9649$, $r = 0.0048$ for $\gamma = 5$; $n_{s }=   0.9712$, $r = 0.0036  $ for $\gamma = 10$;  $n_{s }=   0.9752$, $r= 0.0023$ for $\gamma = 20$. This range is sufficient to match all available CMB and DESI results.

\section{Evolving  dark energy}

\subsection{Waterfalls and quintessence}

In the case of the waterfall with  $\gamma\approx1$, discussed in Sections \ref{sec:examplees} and \ref{sec:wide}, the inflaton potential effectively disappears for $\varphi<\varphi_c$, as shown in Figs. \ref{PotT} and \ref{PotTw}. Indeed, in general, for $\varphi_c-\varphi\gg {\Delta\varphi}$, we can expand the potential in \eqn{cases} as follows
\begin{align}
    V={\frac{V_{\rm original}}{1+\gamma}}\left(1-\gamma+2\gamma e^{2(\varphi-\varphi_c)/\Delta\varphi}+\cdots\right).
\end{align}
Note that $\gamma=1$ is a simple choice and the shape of the potential resembles the quintessential $\alpha$-attractor, but this is too steep for dark energy. If, however, $\gamma-1\neq 0$, the remaining leading constant factor  $\frac{1-\gamma}{1+\gamma}$ (with $0<\gamma<1$) leaves $V_{\rm original}$, which may play the role of the quintessential potential if it is not too steep.

\begin{figure}[H]
\vskip 0.3cm 
\centering
		 \includegraphics[width=0.7\textwidth]{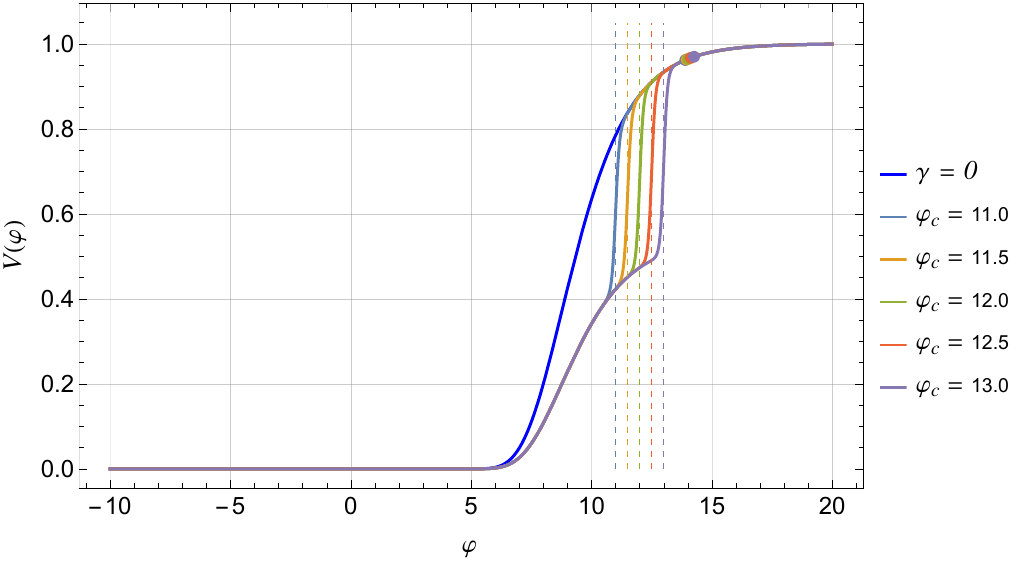}
        \caption{The blue curve describes the original quintessential potential, with $\gamma=0$. 
       This waterfall modulation is the case of the weak waterfall regime, as explained in Sec. \rf{eval}, which temporarily steepens the potential and reduces the number of e-folds accumulated before the eventual end of inflation, thereby increasing $n_s$.}
        \label{PotQ}
\end{figure}

Instead of T- and E-models such as those in \eqn{exampleT} and \eqn{exampleE}, we now consider one of the quintessential 
potentials proposed and studied in  \cite{Dimopoulos:2017zvq,Akrami:2017cir,Zhumabek:2023wka}
\begin{equation}
    V_{\rm quint}( \vp )
        =  M^2 e^{-2g}  \Bigl(e^{g \bigl( \tanh \frac{\vp}{\sqrt{6\alpha}} +1\bigr)}-1\Bigr) \ .
    \label{eq:explin_potential}
\end{equation}
We  now make a waterfall-modulated version of it
\be
V_{\rm {wf}}^{\rm quint}= {1\over 1+\gamma} V_{\rm quint}( \vp )\Big  ( 1+\gamma \tanh {\varphi-\varphi_c\over \Delta \varphi} \Big)\,.
\label{quint}\ee
\begin{figure}[H]
\vskip 0.3cm 
\centering
		 \includegraphics[width=0.85\textwidth]{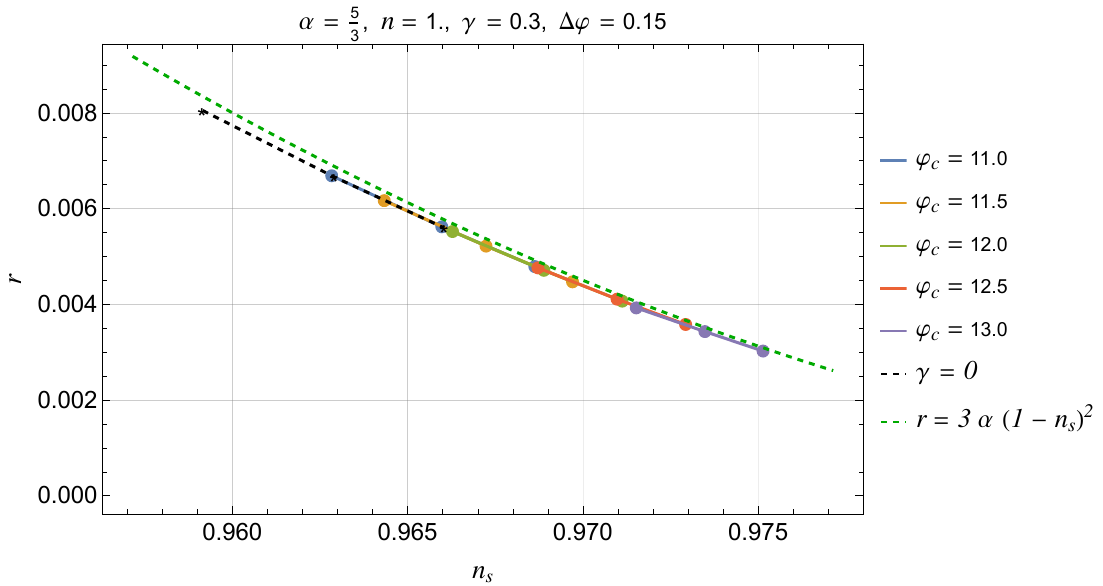}
        \caption{\footnotesize The values of $(n_s, r)$ for waterfall-modulated quintessential models.  }
        \label{nsrwQ}
\end{figure}
\noindent We study the following set of parameters:
\be
g=127.5, \quad \a=5/3, \quad \gamma=0.3, \quad \Delta \vp = 0.15\,.
\ee
 The $(r,n_s)$ results are presented in Fig. \ref{nsrwQ}.
The waterfall-modulated quintessential potentials are shown in Fig.~\ref{PotQ}.

\subsection{More on relation to $\alpha$-attractors}

Now we will return to the derivation of the expression for the step function in the context of $\alpha$-attractors, \rf{tnu}-\rf{nuup}, and, for simplicity, write these equations for real $T =e^{-\sqrt {2  \over 3 \alpha} \varphi }$:
\be\label{tnu1}
V= {V^{T, E}\over 1+\gamma}  \cdot  \left(1+ \gamma\  {T_{c}^\nu -T^\nu \over  T_{c}^\nu +T^\nu }\right),
\ee
This yields
\be
V= {V^{T, E}\over 1+\gamma}   \cdot  \Big[1+ \gamma\ \tanh  {\varphi-\varphi_c\over \Delta \varphi}\Big]  \ ,   \qquad \Delta \vp   = {\sqrt{6\a}\over \nu} \  ,
\ee
or, equivalently, 
\be\label{nuup1}
V= {V^{T, E}\over 1+\gamma}  \cdot  \left[1+ \gamma\ \tanh { \nu(\varphi-\varphi_c)\over  \sqrt{6\alpha}}\right] \ .
\ee
These considerations imply that if one wants to consider potentials with $\Delta \vp \ll 1$, one must use the potentials \rf{tnu1} with very large values of $\nu$.  Therefore, it is important to find out how small $\Delta \vp$ and how large $\nu$ should be to achieve a significant increase of $n_{s}$.

\begin{figure}[H]
\vskip 0.3cm 
\centering
		 \includegraphics[width=0.58\textwidth]{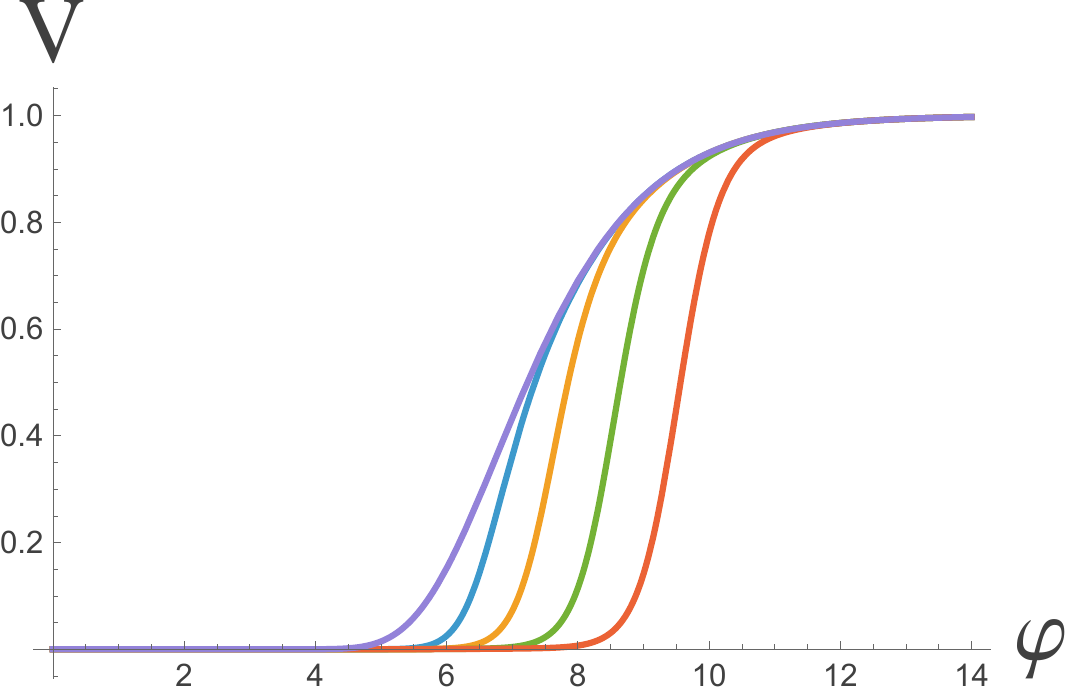}
        \caption{The potential \rf{nuup1} for $\alpha = 1$ and $\gamma = 0.995$ for $\vp_{c} = 6.5$ (blue), 7.5 (yellow), 8.5 (green), and 9.5 (red). The upper (purple) curve shows the unmodulated potential $V_{\rm quint}$  \rf{eq:explin_potential}.}
        \label{verywide}
\end{figure}

Here we consider the quintessential potential \rf{eq:explin_potential} modulated by the function $ 1+\gamma \tanh { 4(\varphi-\varphi_c)\over  \sqrt{6\alpha}} $ \be\label{nuup1}
V= {1\over 1+\gamma}  V_{\rm quint}( \vp ) \cdot  \left[1+\gamma  \tanh { 4(\varphi-\varphi_c)\over  \sqrt{6\alpha } }\right] \ . 
\ee
We will now study it for the particular choice $\alpha = 1$ and $\gamma = 0.995$. This corresponds to $\Delta \vp = 0.612$, which is much wider than the waterfalls in all previously considered examples. The potential is shown in Fig. \ref{verywide} for several different values of $\vp_{c}$.

In this model with $g=127.5$ and $N_*=55$, we find that for $\vp_c=6.5$, $n_s=0.9641$, $r=0.0038$, for $\vp_c=7.5$, $n_s=0.9664$, $r=0.0033$, for $\vp_c=8.5$, $n_s=0.9710$, $r=0.0024$, and for $\vp_c=9.5$, $n_s=0.9745$, $r=0.0013$.

Thus, we have found that in this class of wide waterfall-modulated quintessential potentials with $\nu = 4$ and $\Delta \vp = 0.612$ one can have $n_s$ in the range from  0.9641 to 0.9745 for $N_*=55$. We will study the dark energy aspects of these models separately \cite{Renata2}.

\section{Summary}
The idea of a premature end of inflation was around for a while, starting with hybrid inflation 
\cite{Linde:1991km}, \cite{Linde:1993cn}  and geometric destabilization  \cite{Renaux-Petel:2017dia}. The waterfall destabilization of the normal course of inflation was also studied in the context of hybrid  $\alpha$-attractor models   \cite{Kallosh:2022ggf,Braglia:2022phb,Ye:2022efx}. In all these cases, destabilization was triggered by a second field. 

Recently, it was shown in \cite{Zhang:2026ivx,Yuan:2026xcg} that the effect of destabilization of the normal course of inflation can be simulated in single-field inflationary models by the insertion of the traditional step \cite{Adams:2001vc} in the potential. If the step is of a waterfall type, like in hybrid inflation, the premature end of inflation can increase the effective $N_c$ in the slow-roll parameters in \eqn{src}. This is a desirable feature for the $\a$-attractor potentials in view of the ACT data. On the other hand, if the step flattens the potential, it can decrease the effective $N_c$ and bring monomial inflationary models $\vp^{\beta}$ with $\beta < 1$ closer to the sweet spot of the data.

In this paper, we studied the effect of waterfalls on $\a$-attractor models. We found that the wide and small waterfalls can increase $n_s$ in the single-field $\a$-attractor models up to $n_s\approx 0.975$, in agreement with the evaluation formula for $\a$-attractors we gave in \eqn{we1}, thereby supporting the results in \cite{Zhang:2026ivx}.  We have also found that instantaneous sharp waterfalls can increase $n_s$ up to $n_s\to 1$, confirming the evaluation formula in \eqn{st1} which exhibits the same enhancement mechanism as the hybrid $\alpha$-attractor result in  \cite{Kallosh:2022ggf} and supporting 
single-field $\a$-attractor models studied in \cite{Yuan:2026xcg}.

We have proposed various waterfall insertions into single-field models that are not of the traditional form studied in
\cite{Adams:2001vc}, more recently in \cite{Zhang:2026ivx} and \cite{Yuan:2026xcg}, and here in Secs.~\ref{Sec:2}--\ref{Sec:4}.  The additive uplift constructions introduced in Sec.~\ref{Sec:5}  provide single-field realizations of the second mechanism familiar from hybrid $\alpha$-attractors, namely, an increase of the potential relative to its slope and curvature. 

In conclusion,  we have found that by changing the height, width,  positions, and other features of waterfall insertions, one can densely populate the $\a$-attractor curve $ r\simeq 3\a(1-n_s)^2$ and move continuously from values of $(n_s,r)$ in models without waterfalls to the maximal value of $n_s \simeq 1$.  We have presented an updated forecast for LiteBIRD in Fig. \ref{Flux2} where we show the corresponding descending
curves for  $(n_s,r)$.
Future cosmological experiments  will support or falsify this class of waterfall-modulated $\a$-attractor models.

 We have also investigated the waterfall modulation of the quintessential $\alpha$-attractor models~\cite{Dimopoulos:2017zvq,Akrami:2017cir,Zhumabek:2023wka,Jing:2026ymp} taking into account the effect of the waterfalls and uplift of $n_s$ in these models. Here again, by changing the waterfall features and their positions, one can increase $n_s$ compared to the original models.  Detailed studies of dynamical dark energy in the waterfall-modulated quintessential $\a$-attractor models will be presented in  \cite{Renata2}.  

 \

{\bf Acknowledgments:} We are grateful to  A. Chudaykin, M. Ivanov, and O.H.E. Philcox for collaboration on the recent investigation \cite{Chudaykin:2026amr}, which stimulated the current study, and to Y. Akrami, G. Alestas, and M. Shmakova for discussions on dark energy issues. 
The work of RK and AL is supported by the Leinweber Institute for Theoretical Physics at Stanford and by NSF Grant PHY-2310429. YY is supported by IBS under the project code, IBS-R018-Y3-2026-a00.

\appendix
\section{Evaluation of $N_c$}\label{App:eval}

The expression for $N_c$ was evaluated in \cite{Kallosh:2022ggf} for T-models and presented here in eqs. \rf{h1}, \rf{h2}. Here, we derive the evaluation of $N_c$ for the single-field waterfall models studied in this paper. The results for strong waterfalls are presented in eqs. \rf{st1}, \rf{st2}, \rf{st3}, and those for weak waterfalls in eqs. \rf{we1}, \rf{we2}. In the following, the endpoint defined by $\epsilon_V=1$ is used only as a slow-roll estimate. For sufficiently sharp transitions, the physical end of inflation should instead be determined from $\epsilon_H=1$ using the full background evolution.

To derive the analytic estimate of the effective number of e-folds $N_c$, we write the waterfall potential as
\begin{align}
V_{\rm wf}(\varphi)=V_0(\varphi)F(\varphi),\qquad F(\varphi)=1+\gamma\tanh x,\qquad x\equiv\frac{\varphi-\varphi_c}{\Delta\varphi},
\end{align}
where $V_0$ denotes the corresponding original T- or E-model potential, and the irrelevant overall normalization of $F$ has been omitted. It is useful to introduce
\begin{align}
\lambda_0(\varphi)\equiv\frac{V_0'(\varphi)}{V_0(\varphi)},\qquad s(\varphi)\equiv\frac{F'(\varphi)}{F(\varphi)},\qquad R(\varphi)\equiv\frac{s(\varphi)}{\lambda_0(\varphi)}.
\end{align}
Let $\varphi_e$ and $\varphi_{e,{\rm wf}}$ denote the end points determined by $\epsilon_V=1$ for $V_0$ and $V_{\rm wf}$, respectively. At a fixed horizon-exit value $\varphi_*$, the slow-roll e-fold difference is
\begin{align}
\Delta N=\int_{\varphi_e}^{\varphi_{e,{\rm wf}}}\frac{d\varphi}{\lambda_0(\varphi)}+\int_{\varphi_{e,{\rm wf}}}^{\varphi_*}\frac{R(\varphi)}{1+R(\varphi)}\frac{d\varphi}{\lambda_0(\varphi)}.
\end{align}
The first term accounts for the change in the endpoint, whereas the second term describes the reduction in e-folds due to the increased slope around the step.

We now approximate the region in which $R\gtrsim1$ by a hard window. In practice, this amounts to approximating $R/(1+R)\simeq1$ in the region where $R\gtrsim1$ and neglecting it elsewhere. On the inflationary plateau, the logarithmic slopes of the T- and E-model potentials can be written in the common form
\begin{align}
\lambda_0(\varphi)\simeq b_Xnk\,e^{-k\varphi},\qquad k\equiv\sqrt{\frac{2}{3\alpha}},\qquad b_T=4,\qquad b_E=2,
\end{align}
where $X=T,E$. Defining
\begin{align}
\kappa\equiv k\Delta\varphi,\qquad \lambda_c\equiv b_Xnk\,e^{-k\varphi_c},
\end{align}
the logarithmic slope of the step has the asymptotic forms
\begin{align}
s(x)\simeq\frac{4\gamma}{(1+\gamma)\Delta\varphi}e^{-2x}\quad(x\gg1),\qquad s(x)\simeq\frac{4\gamma}{(1-\gamma)\Delta\varphi}e^{2x}\quad(x\ll-1),
\end{align}
where the lower-tail expression applies for $0<\gamma<1$. The upper and lower crossover points, defined by $R(\varphi_\pm)=1$, are therefore
\begin{align}
\varphi_\pm=\varphi_c+\Delta\varphi\,x_\pm,\qquad x_+=\frac{\log Q_+}{2-\kappa},\qquad x_-=-\frac{\log Q_-}{2+\kappa},
\end{align}
with
\begin{align}
Q_\pm\equiv\frac{4\gamma}{(1\pm\gamma)\Delta\varphi\,\lambda_c}.
\end{align}
The lower boundary of the hard-window contribution is
\begin{align}
\varphi_L\equiv\max\!\left(\varphi_{e,{\rm wf}},\varphi_-\right).
\end{align}
The e-fold difference is then approximated by
\begin{align}
\Delta N\simeq\int_{\varphi_e}^{\varphi_{e,{\rm wf}}}\frac{d\varphi}{\lambda_0(\varphi)}+\int_{\varphi_L}^{\varphi_+}\frac{d\varphi}{\lambda_0(\varphi)}.
\end{align}
Using the exponential approximation for $\lambda_0$, this becomes
\begin{align}
\Delta N\simeq C_X\left[e^{k\varphi_{e,{\rm wf}}}-e^{k\varphi_e}+e^{k\varphi_+}-e^{k\varphi_L}\right],\qquad C_X\equiv\frac{1}{b_Xnk^2}=\frac{3\alpha}{2b_Xn},
\end{align}
where
\begin{align}
C_T=\frac{3\alpha}{8n},\qquad C_E=\frac{3\alpha}{4n}.
\end{align}

If the inflaton passes through the step-dominated region before inflation ends, $\varphi_{e,{\rm wf}}\leq\varphi_-$ and $\varphi_{e,{\rm wf}}\simeq\varphi_e$. The result reduces to
\begin{align}
\Delta N^{\rm weak}\simeq C_Xe^{k\varphi_c}\left[Q_+^{\frac{\kappa}{2-\kappa}}-Q_-^{-\frac{\kappa}{2+\kappa}}\right].\label{weak N}
\end{align}
If inflation terminates before the lower crossover is reached, $\varphi_{e,{\rm wf}}>\varphi_-$, the dependence on $\varphi_{e,{\rm wf}}$ cancels and one finds
\begin{align}
\Delta N^{\rm strong}\simeq C_X\left[e^{k\varphi_c}Q_+^{\frac{\kappa}{2-\kappa}}-e^{k\varphi_e}\right].\label{strong N}
\end{align}
For a waterfall located sufficiently high on the plateau, the last term is subleading, and we obtain
\begin{align}
\Delta N^{\rm strong}\simeq C_Xe^{k\varphi_c}Q_+^{\frac{\kappa}{2-\kappa}}.\label{strong N2}
\end{align}
For $\gamma=1$, the lower-tail expression and $Q_-$ are not used; the strong-waterfall expression above directly applies.

The effective number of e-folds entering the attractor expressions is therefore
\begin{align}
N_c \simeq N_*+\Delta N,
\end{align}
which gives the slow-roll parameters in \eqn{src}.

\bibliographystyle{JHEP}
\bibliography{lindekalloshrefs}
\end{document}